\documentclass[twocolumn,preprint]{aastex631}
\usepackage{amsmath}

\newcommand{\beq}{\begin{equation}}
\newcommand{\eeq}{\end{equation}}
\newcommand{\beqa}{\begin{eqnarray}}
\newcommand{\eeqa}{\end{eqnarray}}

\def\stacksymbols #1#2#3#4{\def\theguybelow{#2}
        \def\verticalposition{\lower#3pt}
        \def\spacingwithinsymbol{\baselineskip0pt\lineskip#4pt}
        \mathrel{\mathpalette\intermediary#1}}
\def\intermediary #1#2{\verticalposition\vbox{\spacingwithinsymbol
        \everycr={}\tabskip0pt
        \halign{$\mathsurround0pt#1\hfil##\hfil$\crcr#2\crcr
                \theguybelow\crcr}}}

\begin{document}
\title{PAH Spectroscopy from 1-5 $\mu$\lowercase{m}.  }

\author{L. J. Allamandola, C. Boersma, T. J. Lee, J. D. Bregman, and P. Temi}
\affiliation{NASA Ames Research Center, MS 245-6, Moffett Field, CA 94035-0001, USA; allamandola@sbcglobal.net}
\affiliation{Received 2021 June 5; revised 2021 July 18; accepted 2021 July 26; published 2021 August 24}

%
%
%
%
%

\begin{abstract}

  The PAH model predicts many weak emission features in the 1–5 $\mu$m region that can resolve significant questions that it has faced since its inception in the mid-80s. These features contain fundamental information about the PAH population that is inaccessible via the much stronger PAH bands in the 5–20 $\mu$m region. Apart from the 3.3 $\mu$m band and plateau, PAH spectroscopy across most of the 1–5 $\mu$m region has been unexplored due to its low intrinsic intensity. ISO and Akari covered some of this wavelength range, but lacked the combined sensitivity and resolution to measure the predicted bands with sufficient fidelity. The spectroscopic capabilities of the NIRSpec instrument on board JWST will make it possible to measure and fully characterize many of the PAH features expected in this region. These include the fundamental, overtone and combination C–D and C$\equiv$N stretching bands of deuterated PAHs, cyano-PAHs (PAH-C$\equiv$ N), and the overtones and combinations of the strong PAH bands that dominate the 5–20 $\mu$m region. These bands will reveal the amount of D tied up in PAHs, the PAH D/H ratio, the D distribution between PAH aliphatic and aromatic subcomponents, and delineate key stages in PAH formation and evolution on an object-by-object basis and within extended objects. If cyano-PAHs are present, these bands will also reveal the amount of cyano groups tied up in PAHs, determine the N/C ratio within that PAH subset, and distinguish between the bands near 4.5 $\mu$m that arise from CD versus C$\equiv$N.\\
\end{abstract}

\vskip1.5cm


\section{Introduction}
\noindent
The infrared (IR) spectra from many galactic and extragalactic objects are dominated by emission features that peak near 3.3, 6.2, 7.7, 8.6, and 11.2 $\mu$m (e.g., Peeters et al. 2002; Tielens 2008; Li 2020, and references therein). These spectra are generally attributed to IR fluorescent emission from highly vibrationally excited polycyclic aromatic hydrocarbon (PAH) molecules pumped by UV photons. Since PAHs are molecules (nano-sized), not particles, they play important astrophysical and astrochemical roles distinct from those of grains. Their unique properties, coupled with their spectroscopic response to changing conditions and their ability to convert UV to IR radiation, makes them important players in, and powerful probes of, a wide range of astronomical environments and object types.

PAHs account for roughly 15\% of the interstellar C and their emission is responsible for up to 20\% of the IR power of the Milky Way and star-forming galaxies (Smith et al. 2007; Tielens 2008; Li 2020). They dominate cloud cooling and are efficient catalysts for $H_2$ formation (Thrower et al. 2012; Boschman et al. 2015, and refs therein). By providing photoelectrons, they influence the thermal budget and chemistry of the interstellar medium (ISM) and control the heating of the gas in the diffuse ISM and surface layers of protoplanetary disks (Bakes \& Tielens 1994; Weingartner \& Draine 2001; Kamp \& Dullemond 2004). As an important electron sink, PAHs dominate the ionization balance in molecular clouds and thus influence ion-molecule and ambipolar diffusion processes that set the stage for star formation by coupling magnetic fields to the gas (Verstraete 2011). Through their influence on the forces supporting clouds against gravitational collapse, PAHs affect the process of star formation itself and the phase structure of the ISM (Tielens 2008).

Despite the continuous advancements made with the interstellar PAH model, pertinent questions and uncertainties remain and new ones continue to arise. These uncertainties severely limit our understanding of how PAHs respond to and affect physical and chemical conditions in different astronomical environments. Many of these uncertainties can be overcome by tapping into the unique molecular information contained in the 1–5 $\mu$m region of PAH emission spectra. For example, while the recent discoveries of cyanonaphthalene (C10 H7-C$\equiv$N) in TMC-1 and fullerenes (C60) in TC-1 signify the first detection of specific members of the PAH family in space (e.g., Cami et al. 2010; Sellgren et al. 2010; McGuire et al. 2021), they also raise very significant, fundamental questions regarding PAH formation, evolution, and excitation. Moreover, if cyano-PAHs prove to be important members of the interstellar PAH population, their characteristic C$\equiv$N stretching band near 4.5 $\mu$m may confound the attribution of the features in the same region to deuterated PAHs (Peeters et al. 2004; Doney et al. 2016). Furthermore, the PAH model has advanced to such a degree that it now allows the treatment of anharmonicity in determining emission band positions in the 1–5 $\mu$m region (e.g., Mackie et al. 2018b; Maltseva et al. 2015; Chen et al. 2019).

\begin{figure*}[ht!]
  \begin{center}
    \includegraphics[width=16cm]{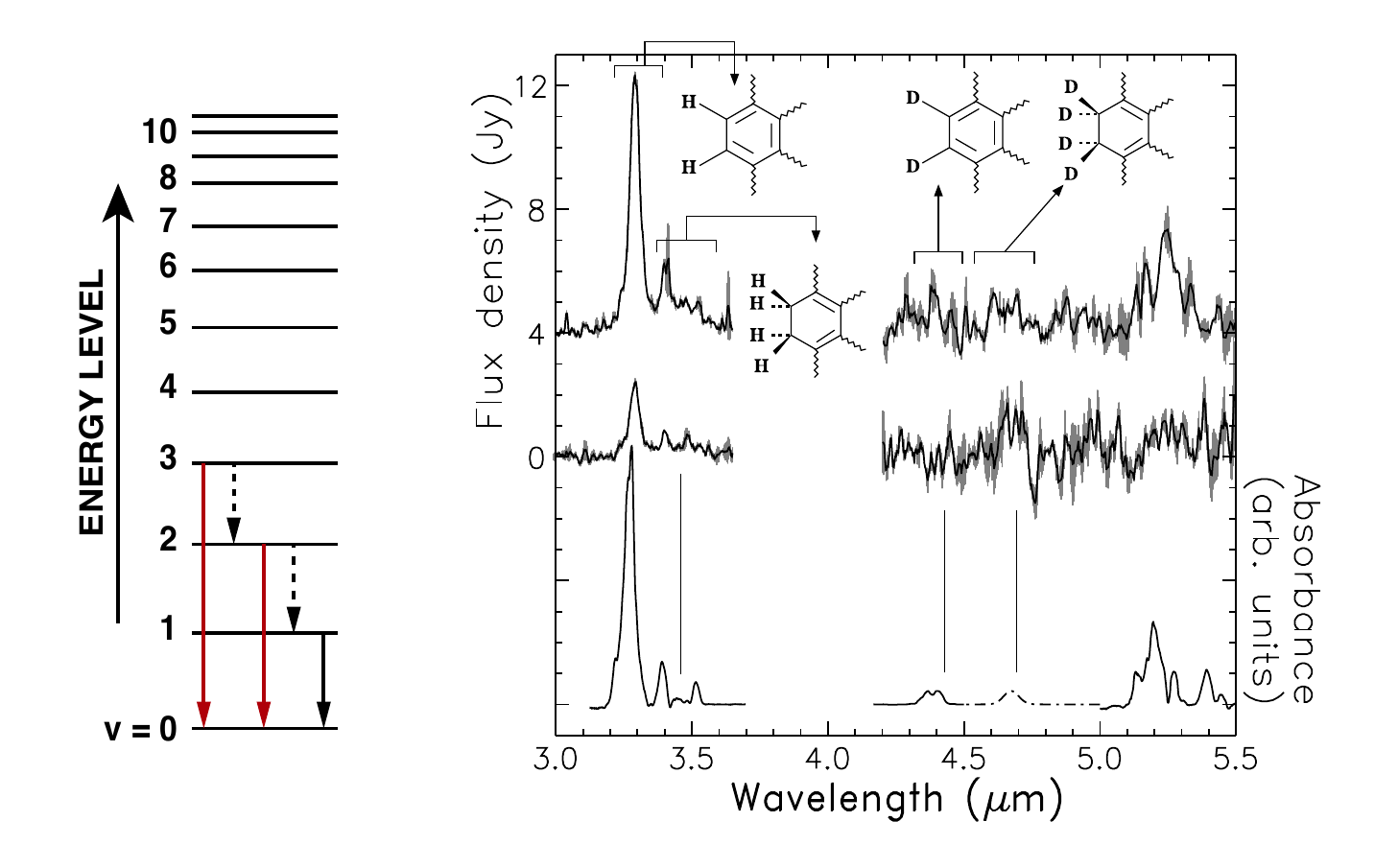}

    \caption{
      Left panel: Emission transitions in an anharmonic vibrational energy level diagram for a molecular vibrational mode. Numbers indicate quantized vibrational levels, v. The solid black arrow indicates the fundamental transition; dashed black arrows, hot bands; and red arrows, overtones. Right panel: Comparison of the 3.0–5.5 $\mu$m ISO-SWS spectrum of the Orion Bar (top trace) and M17 (middle trace) to the laboratory absorption spectrum of a partially deuterated, fully aromatic PAH mixture (lower trace, solid line) and the computed absorption spectrum of a partially aliphatic, deuterated PAH (lower trace, dashed line). Chemical structures of PAH edge rings illustrating these bonding types are shown at the top of the figure. Stretching of the CH and CD bonds in the chemical structures shown give rise to the bracketed spectral features. The dashed lines and elongated triangles in the structures indicate the H and D atoms that are below or above the molecular plane, respectively, and the wiggly lines represent bonds to the hexagonal carbon skeleton (Hudgins et al. 2004). The CD stretches produce the bands peaking near 4.4 and 4.7 $\mu$m, while the corresponding CH stretches produce bands in the 3 $\mu$m region. The 5.2 $\mu$m feature is produced by overtones and combinations of CH-bending vibrations (e.g., Boersma et al. 2009). Figure adapted from Peeters et al. (2004). See Hudgins et al. (2004) for discussion of the experimental and computational data shown in the lower trace.
    }
    \label{fg:levels}
  \end{center}
\end{figure*}

Due to its low intensity (e.g., Li \& Draine 2001), the 1–5 $\mu$m region has been largely unexplored for objects that show the strong 5–15 $\mu$m PAH bands. This will change with the launch of the James Webb Space Telescope (JWST). The spectroscopic capabilities of its NIRSpec instrument will make it possible to measure many of the PAH features expected in this region. Except for the 2.8–3.7 $\mu$m section, which is dominated by emission from the PAH CH fundamental stretch and its satellite features, the rest of the 1–5 $\mu$m region is free of strong PAH fundamental bands. As described in this work, the PAH emission model predicts that many important, but weaker, features are spread across this entire spectral range. From their band positions, profiles, and relative intensities one can glean information about the interstellar PAH population and local astrophysical conditions that cannot be obtained from the prominent 5–15 $\mu$m features. Indeed, there have already been some tantalizing hints of the predicted PAH features in this region, including the 1.68 $\mu$m band (Geballe et al. 1994) and several features near 4.5 $\mu$m (Peeters et al. 2004; Doney et al. 2016).

This Letter is organized as follows. Section 2 summarizes aspects of emission spectroscopy that are relevant to this work and describes the fundamental and overtone bands of PAHs, deuterated PAHs, and cyano-PAHs that fall in the 1–5 $\mu$m region. Section 3 discusses the astronomical implications of this new information. Section 4 summarizes and concludes the Letter.

\vskip-1cm
\section{
  Interstellar PAH spectroscopy from 1-5 $\mu$\lowercase{m}
 }

Interstellar PAH IR spectra are emission spectra from molecules that have become highly vibrationally excited primarily by absorbing UV photons. Under the collision-free conditions of the ISM, once excited, the main relaxation channel available is through the stepwise emission of IR photons at wavelengths corresponding to the fundamental vibrational modes of the molecule, i.e., CH stretching and bending modes, CC stretches and bends, etc. These are the most intense modes and, except for the 3.3 $\mu$m CH stretch, they all fall between 5 and 15 $\mu$m (e.g., Li 2020; Peeters et al. 2002). However, these are not the only vibrational transitions through which an excited molecule can relax. The left panel in Figure 1 shows an energy level diagram that is typical for a molecular vibrational mode. The familiar PAH features are dominated by fundamental transitions, i.e., those that emit from the first vibrational level (v=1$\rightarrow$0, solid black arrow in Figure 1). In addition to the intense fundamental transition, if sufficiently populated, vibrationally excited molecules can also emit from higher levels within a particular vibrational mode via hot bands such as the v=2$\rightarrow$1, and v=3$\rightarrow$2 transitions (dashed black arrows in Figure 1), overtone transitions such as v=2$\rightarrow$0, and v=3$\rightarrow$0 transitions (red arrows in Figure 1), and combination bands that involve coupling between vibrational levels from different vibrational modes. Combination bands have been described in Appendix A. The Einstein A values of hot bands are similar to those of the fundamentals whereas overtone and combination band intensities are significantly weaker.

\begin{table}[ht]
  \small
  \caption{Wavelength Regions of PAH C–D and C$\equiv$N Stretching Bands}
  \begin{center}
  \begin{tabular}{cccc}
    \hline
    \hline
    Vibration          & $\Delta$v=1$\rightarrow$0 & $\Delta$v=2$\rightarrow$0 & $\Delta$v=3$\rightarrow$0 \\
                       & $\lambda$ ($\mu$m)        & $\lambda$ ($\mu$m)        & $\lambda$ ($\mu$m)        \\
    \hline
    PAH                & 4.30 - 4.50$^a$           & 2.16 - 2.21$^c$           & 1.45 - 1.48$^c$           \\
    CD stretch         &                           &                           &                           \\
    \hline
    DPAH               & 4.54 - 4.75$^a$           & 2.32 - 2.37$^c$           & 1.57 - 1.60$^c$           \\
    CD stretch         &                           &                           &                           \\
    \hline
    PAH-nitrile        & 4.46 - 4.50$^b$           & 2.26 - 2.27$^c$           & 1.518 - 1.523$^c$         \\
    C$\equiv$N stretch &                           &                           &                           \\
    \hline
  \end{tabular}
\end{center}
  {\bf Notes:} Fundamental ($\Delta$v=1$\rightarrow$0) wavelength ranges of the aromatic CD stretch in PADs, the aliphatic CD stretch in DPAHs, the C$\equiv$N stretch for cyano-PAHs, and the anharmonic wavelengths for their $\Delta$v=2$\rightarrow$0, and $\Delta$v=3$\rightarrow$0 overtone bands determined using the method described in Appendix B.

  $^a$~Hudgins et al., 2004.\\
  $^b$~Silverstein and Bassler, 1967.\\
  $^c$~See Appendix B
  \label{tbl:hst}
\end{table}

The positions and relative intensities of these weak features are the key to accessing the fundamental information about PAH structure, composition, charge, and size that is entangled in the 5–15 $\mu$m spectra. To gain this access requires knowing the anharmonic vibrational energy levels for astronomically relevant PAHs. Fortunately, although this information is currently very limited, the field is opening up. A recent series of breakthrough experimental and computational papers has shown that anharmonicity and Fermi-type resonances play key roles in determining the spectra of cold ($\sim$10 K), gas-phase PAHs in the 3 $\mu$m region (Mackie et al. 2018a, 2018b; Maltseva et al. 2018, and references therein). This tour de force work showed that Fermi resonances and other interactions between overtone and combination bands with the fundamental CH stretching bands in the 3 $\mu$m region are required to explain the overall profile and intensities of the observed 3.3 $\mu$m emission feature, its broad underlying plateau, and substructure. Focusing on the 1.6–1.7 $\mu$m region, Chen et al. (2019) examined the effects of Fermi and Darling–Dennison resonances exclusively on anharmonic combination bands. Fermi and Darling–Dennison resonances involve coupling between two and three different vibrational modes, respectively. Understanding all of these details taken together has been a decades-long challenge in astronomical PAH spectroscopy. Mackie et al. (2018a, 2018b) and Maltseva et al. (2018) have unequivocally shown that anharmonicity plays a key role in determining which overtone and combination bands are involved in the Fermi resonance polyads and symmetry interactions that drive significant band intensity sharing and thereby determine overall spectral structure of the observed 3.3 $\mu$m feature. The method described by Mackie et al. (2018a), used here to determine the anharmonic overtone band positions listed in Table 1 and shown in Figure 2, has been summarized in Appendix B. A discussion as to why combination bands and overtone bands are treated differently in this Letter is given at the start of Appendix A, with specific details given in Appendix B.

\begin{figure*}[ht!]
  \begin{center}
    \hskip-0.0cm
    \includegraphics[width=13.8cm]{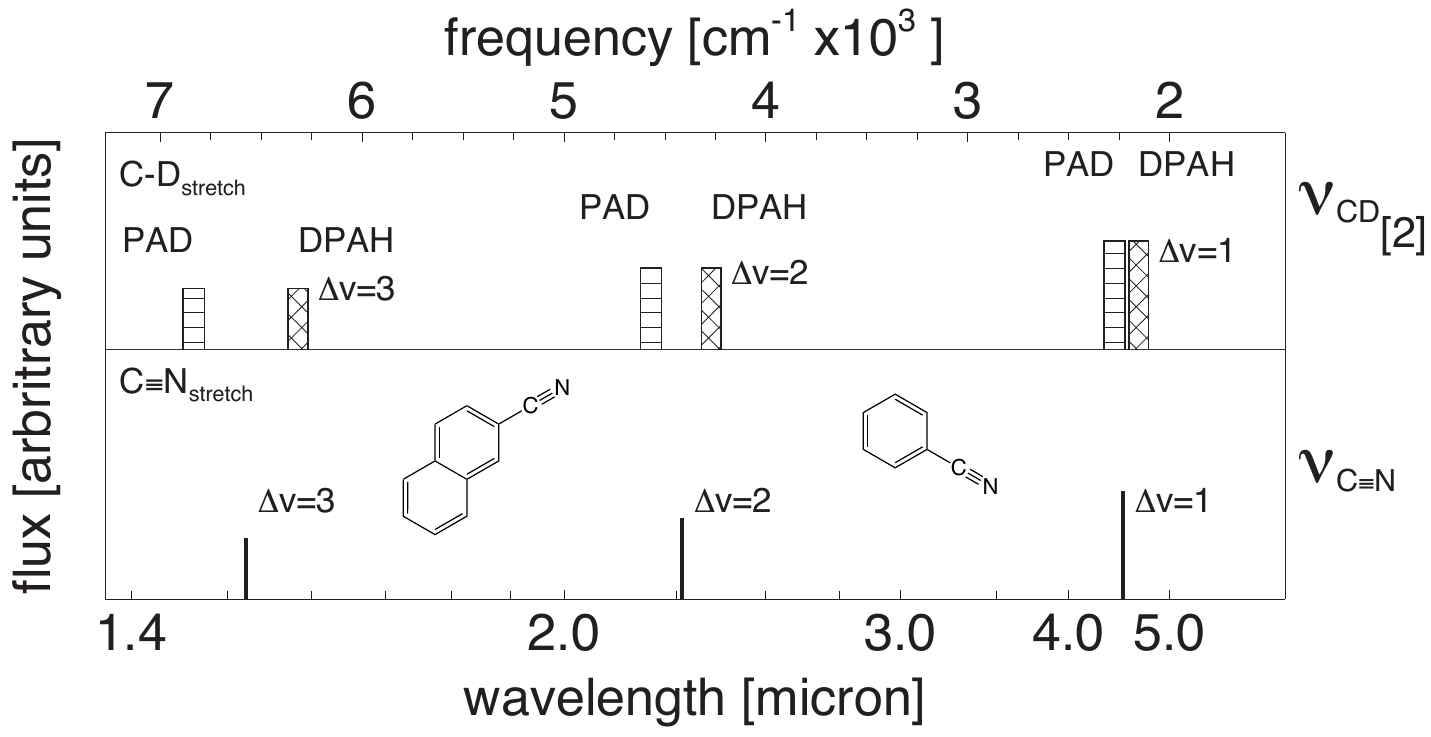}
    \caption{Fundamental and overtone wavelength ranges for the aromatic (horizontal lined) and aliphatic (cross–hatched) C–D stretch in deuterated PAHs (top) and the C$\equiv$N stretch in cyano-PAHs (bottom). The method used to determine overtone band positions is described in Appendix B. Band intensities are shown falling 25\% at each step. The structure of benzonitrile (right) and 3-cyanonaphthalene (left) are shown in the bottom panel.
      .}
    \label{fg:bands}
  \end{center}
\end{figure*}

\subsection{The C-D stretching band in deuterated PAHs and the C$\equiv$N stretching band in Cyano-PAHs}
\noindent
{\it The C-D stretch}: The fundamental CD stretching bands for all deuterated PAHs and their first and second anharmonic overtone bands fall only in the 1–5 $\mu$m region. These are shown in Figures 1 and 2 and listed in Table 1. The right panel of Figure 1 shows that the fundamental C–D stretch for deuterated PAHs (PADs), in which a D atom simply replaces a peripheral H atom, falls roughly between 4.30 and 4.51 $\mu$m. The figure also shows that the fundamental CD stretch for deuterated PAHs (DPAHs), in which the D atom adds to a peripheral carbon atom already containing an H atom, falls between 4.54 and 4.75 $\mu$m (Hudgins et al. 2004). In the former case, PAH planarity and aromaticity are preserved, with the bonds referred to as aromatic C–D bonds. In the latter case, PAH planarity is broken, with the added D atoms and affected C atoms displaced slightly above and below the molecular plane. These are referred to as aliphatic C–D bonds. Aliphatic C–D stretching bonds are also associated with PAHs for which D has replaced H on attached -CH2 and -CH3 functional groups.

\noindent
{\it The C$\equiv$N stretch}: The recent discoveries of radio emission from benzonitrile (C6 H5-C$\equiv$N), two cyanonaphthalenes (1-C10 H7-C$\equiv$N, 2-C10 H7-C$\equiv$N), and other small aromatic species containing -C$\equiv$N in TMC-1 (McGuire et al. 2018, 2021) suggest cyano-PAHs may be an important subset of the interstellar PAH family and thus should be considered here. The fundamental CN stretch in all cyano-PAHs and their first and second anharmonic overtone bands fall only in the 1–5 $\mu$m region. These are shown in Figure 2 and listed in Table 1. The fundamental ($\Delta$v=1$\rightarrow$0) C$\equiv$N stretching bands for all cyano-PAHs falls between 4.46 and 4.50 $\mu$m (e.g., Bellamy 1960; Silverstein \& Bassler 1967).

\subsection{PAH Overtone, Combination, and Difference Bands}

Unfortunately, knowledge of the anharmonicities of the vibrational modes that produce overtone, combination, and difference bands in the 1–5 $\mu$m region for large PAHs is limited. Scaled harmonic wavelengths of the overtone bands, based simply on the peak positions of the canonical interstellar PAH emission bands between 5 and 15 $\mu$m, are shown in Figure 3. These were determined by multiplying the frequencies of the canonical bands by 2, 3, 4, etc. For example, the scaled harmonic, $\Delta$v=2$\rightarrow$0  overtone of the 6.2 $\mu$m (1613 $cm^{-1}$) band is 3.1 $\mu$m (3226 $cm^{-1}$).

While taking anharmonicity into account will slightly redshift the band positions shown from a few hundredths to several tenths of a micron with increasing overtone level, the point of the figure is to illustrate that overtone bands span the 1–5 $\mu$m region. Except for the $\Delta$v=1$\rightarrow$0 fundamental of the CH stretch and $\Delta$v=2$\rightarrow$0 ($\Delta$v=2) overtones of the CH out-of-plane bending bands (CH$_{oop}$ bends) which contribute to the 5.2 and 5.7 $\mu$m emission bands (Boersma et al. 2009), overtone transitions of the PAH 5–15 $\mu$m bands spread across the 1–5 $\mu$m region. In addition to overtone bands, emission from the very large number of PAH combination and difference transitions between states in different vibrational modes also fall across the region. A schematic representation of a few PAH combination bands is shown in Appendix A.

\begin{figure*}[ht!]
  \begin{center}
    \hskip-0.0cm
    \includegraphics[width=13.8cm]{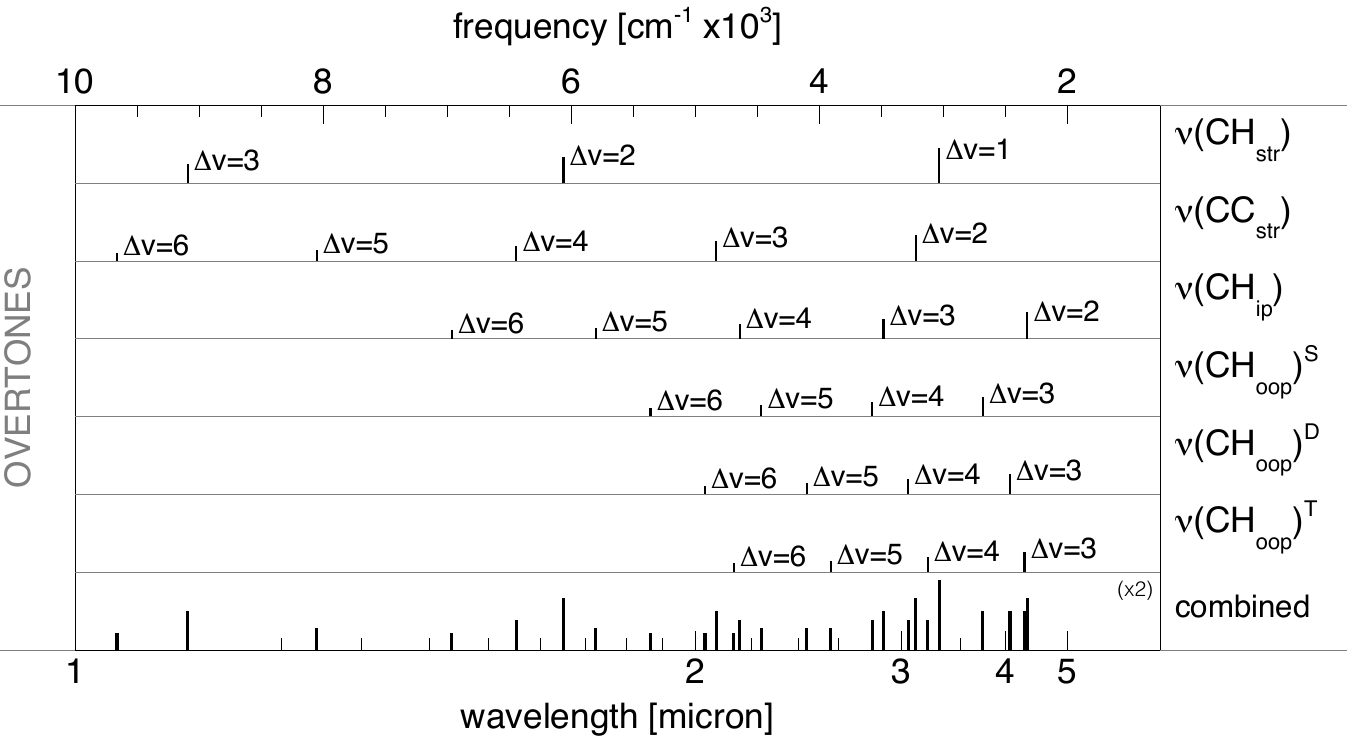}
    \caption{Schematic representation of the scaled harmonic positions of the 1–5 $\mu$m PAH overtone bands, up to $\Delta$v=6, originating from the strong PAH bands between 5 and 15 $\mu$m . These are plotted together in the bottom row (x2). Intensities are shown to fall by 25\% with each step within a vibrational mode progression. Vibrational mode types: str- stretch, ip- in plane CH bend, oop- out of plane CH bend, S- solo hydrogen, D- duo hydrogen, T- trio hydrogen.}
    \label{fg:overtones}
  \end{center}
\end{figure*}

\section{Astronomical Implications}

PAH emission is associated with objects at nearly every stage of the low-mass, stellar lifecycle and PAHs are thought to play key roles in the astrophysical processes associated with many of these object types. The following discussion traces PAH evolution along the stellar lifecycle, focusing on how the 1–5 $\mu$m features can be used to test various hypotheses regarding PAH evolution and the astrophysical processes they influence.

The PAHs that give rise to the familiar 5–15 $\mu$m features are thought to form in carbon-rich stellar envelopes (CSE, Stage 1, Figure 4) as they shine brightly in proto- and planetary nebulae (PPNe, PNe, Stage 2). PAHs passing through the CSE-PPNe-PNe phases are not expected to carry D as it is consumed, not produced in stars (Linsky et al. 2006). However, cyano-PAHs (PAH-C$\equiv$N) are likely to be present because nitrile (C$\equiv$N) and alkyne (C$\equiv$C-H) radicals, PAH building blocks, are both abundant in CSEs (Cernicharo et al. 2000, 2001; Fonfria et al. 2008) and have been shown to be co-spatial in ALMA maps of carbon stars (Agúndez et al. 2017). Cyano-PAHs and alkyne-PAHs are considered key intermediates in PAH growth and destruction reactions in CSEs and PPNe (Hamid et al. 2014; Kaiser et al. 2015; Soliman et al. 2015; Parker \& Kaiser 2017; Santoro et al. 2020). Thus, a mixture of pure PAHs, cyano-PAHs, and PAHs containing nitrogen within the hexagonal network should be important members of the PAH family ejected into the ISM from CSEs and PNe. Consequently, if this hypothesis stands, emission features between 4.46 and 4.51 $\mu$m from PPNe and young PNe should come exclusively from cyano-PAHs, with no PAD blending. After passing through the PNe phase and into the diffuse ISM (Stage 3), the nascent PAH population is subjected to millions of years of UV irradiation, particle bombardment, shocks, and reactions with hydrogen, deuterium, and simple interstellar compounds. Here, D enrichment occurs and only the most robust PAHs survive, i.e., those with more than $\sim$40–50 C atoms, having compact, symmetric structures, perhaps some with and some without N incorporated, and with no side groups such as nitriles. Here, emission features between 4.4 and 4.65 $\mu$m are expected to come exclusively from PADs and DPAHs. As densities increase when molecular clouds and protostellar regions form (Stage 4), PAH processing now includes reactions within and on the surfaces of ice particles and ice-covered grains. Once in the star- and planet-forming disk phase (Stage 5), radiation from the young star drives PAH-ice chemistry that leads to deuterium scrambling and further deuteration, the addition of -C$\equiv$N and other side groups and, quite likely, new PAH formation as well (Bernstein et al. 1999; Sandford et al. 2000; Dworkin et al. 2004; Bernstein et al. 2002; Greenberg et al. 2000). As these ices warm, sublime, and sputter, they introduce an entirely new PAH population into the gas.

\subsection{Deuterated PAHs}

Unlike the cosmically abundant elements C, N, and O, deuterium was only formed immediately following the Big Bang. Since there are no known stellar processes that form D, and D is astrated in stars, its abundance has only decreased since its creation (Reeves et al. 1973 and references therein). The relative reduction of deuterium from its primordial value indicates the fraction of the ISM that has been processed in stars. Thus, measuring both the primordial and current deuterium abundances provides important information about the origin of the Universe and the evolution of our galaxy. However, measurements of the interstellar gas in the Galactic disk reveal a very wide range in the observed D/H ratio that is unlikely to be explained by variable astration rates (Linsky et al. 2006; Draine 2004). The large-scale sequestering of D by PAHs may contribute to the observed variability.

PAHs are thought to become D-enriched exclusively via gas-phase processes in the diffuse ISM and, in dense molecular clouds, via gas-phase processes and dust/ice chemistry. The amount of enrichment is sensitive to object history and local physical conditions (Peeters et al. 2004; Kalvans \& Shmeld 2018; Doney et al. 2016; Onaka et al. 2014; Sandford et al. 2000). Driven by UV radiation and other energetic processes in the diffuse ISM, D constantly exchanges with the H atoms attached to PAHs. Since C–D bonds are slightly more thermodynamically stable than C-H bonds, with time, D enrichment should build (Wiersma et al. 2020).

The PAH D/H ratio is expected to vary from close to zero up to $\sim$0.4 along the stellar lifecycle (Peeters et al. 2004; Onaka et al. 2014; Doney et al. 2016). Unfortunately, it has not yet been possible to put this hypothetical picture of variable PAH deuteration to an observational test because of instrumental limitations and low emission intensity. By comparing the strength of the deuterated PAH emission features between 4.3 and 4.7 $\mu$m (Figure 1) to the already known PAH C-H stretching band intensities between 3.25 and 3.5 $\mu$m along the same line-of-sight, one can constrain the amount of cosmological D sequestered in PAHs for the entire IR-emitting PAH family and directly measure its C–D/C-H ratio. However, as mentioned in Section 2.1, the recent discovery of interstellar cyano-PAHs, molecular species that also have bands in the 4.3–4.8 $\mu$m range, questions assigning emission in this region solely to PADs and DPAHs. Fortunately, as shown in Section 2.1, overtone and combination bands of the CD stretch in deuterated PAHs, if detectable, can be used to distinguish between an origin in cyano-PAHs, PADs, and DPAHs. Further confusing the issue, resonances between PAH overtone and combination bands with the fundamental CD and CN stretching bands will contribute to the overall structure of emission in the 4.5 $\mu$m region much as resonances between the PAH overtone and combination bands with the CH stretch determine the overall structure of emission in the 3.3 $\mu$m region.

Nonetheless, measuring the 4–5 $\mu$m spectra of objects at different stages of the stellar lifecycle has the potential to distinguish between deuterated PAH and cyano-PAH emission in objects for which one of these is dominant and, perhaps most importantly, to assign their relative contributions in objects where both are present.

\begin{figure*}[ht!]
  \begin{center}
    \hskip-0.0cm
    \includegraphics[width=18.0cm]{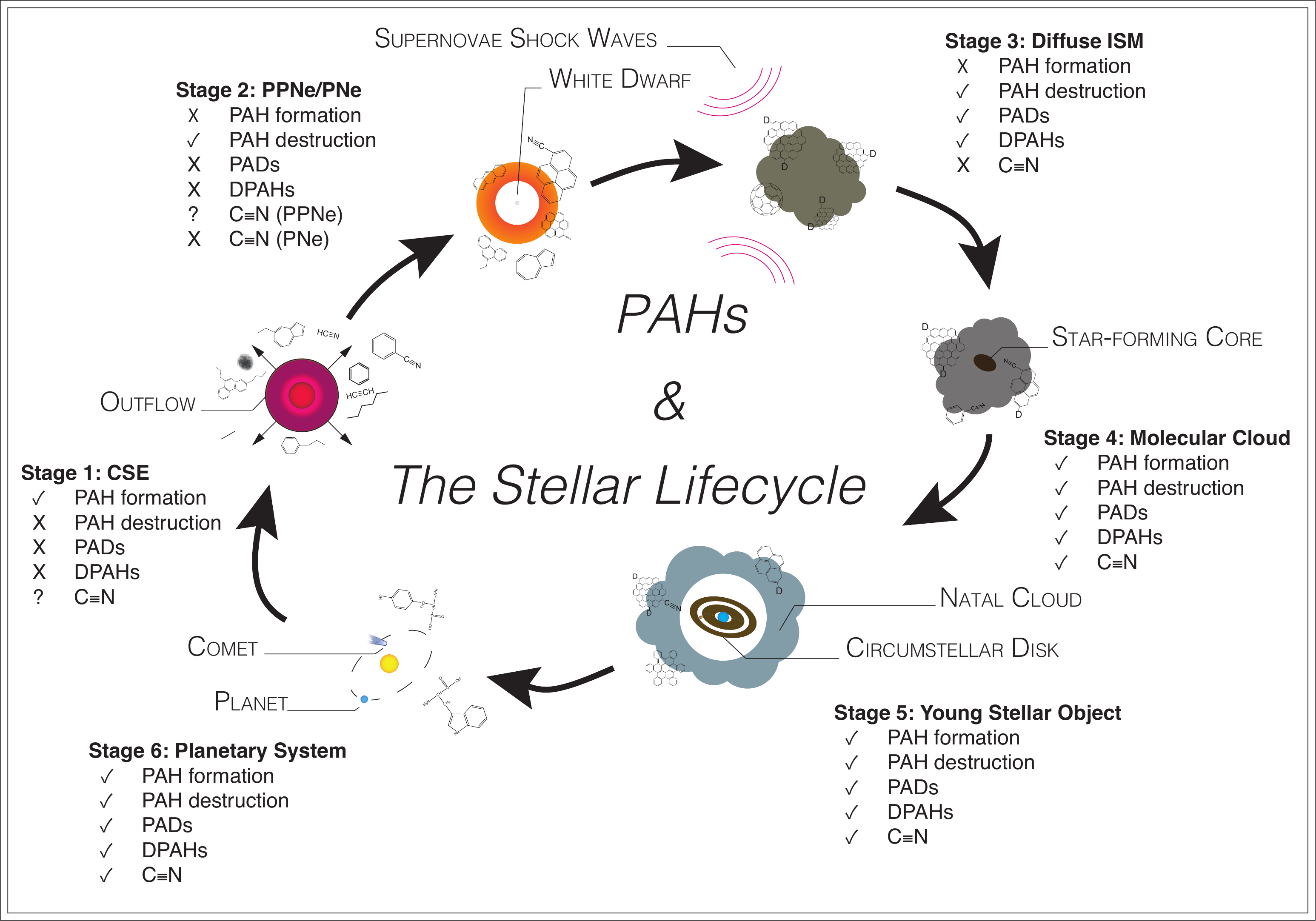}
    \caption{
      PAH populations and processes along the low-mass stellar lifecycle. Check marks indicate the importance of PAH formation and/or destruction during each stage and that IR emission from PADs, DPAHs, and cyano-PAHs is likely to be detected, $\checkmark$ indicates that IR emission from PADs, DPAHs, and cyano-PAHs is not expected. Stage 1 and 2 acronyms CSE and PPNe/PNe represent C-rich stellar envelopes and protoplanetary nebula/planetary nebula, respectively.
    }
    \label{fg:cycle}
  \end{center}
\end{figure*}

\subsection{Cyano-PAHs}

The discovery of radio emission from benzonitrile and two cyanonaphthalenes in the molecular cloud TMC-1 (C$_6$H$_5$-C$\equiv$N; C$_{10}$H$_7$-C$\equiv$N; Figure 3; McGuire et al. 2018, 2021) is a major milestone in the development of the PAH model. As with deuterated PAHs, cyano-PAH abundances are expected to vary as a function of object-type and across extended objects. One of the reasons the discovery of benzonitrile and cyanonaphthalenes in TMC-1 is so remarkable is that it indicates that they were formed in the molecular cloud itself. Benzonitrile and the cyanonaphthalenes are far too small to have survived the rigors of the diffuse ISM, suggesting additional interstellar PAH formation routes and chemistry.

As mentioned above, another reason this discovery is so important is because the fundamental C$\equiv$N and CD stretches in PAHs may blend. However, as Figure 2 demonstrates, measuring the C$\equiv$N and C–D overtones near 2.3 and 1.5 $\mu$m can settle this since they clearly separate. The expected dependence of the presence of C$\equiv$N vs. CD-containing PAHs on the evolutionary stage of the host object also has the potential to break the potential degeneracy of bands between 4.3 and 4.8 $\mu$m for objects where both are present.

By measuring the strength of the cyano-PAH emission features between 4.46 and 4.51 $\mu$m, one can quantify the abundance of cyano groups tied up in PAHs for the entire IR-emitting PAH family and, if present, determine the N/C ratio within the cyano-PAH subset. In addition, measuring the C$\equiv$N fundamental and overtone bands for PAH-nitriles will distinguish between deuterated PAHs and cyano-PAHs in objects where both are present, track the growth and loss of PAH-nitriles through different stages of the stellar lifecycle, and reveal key stages in PAH formation and evolution.

\subsection{PAH Overtone, Combination, and Difference Bands}

Without sufficient spectral resolution, emission from the large number of distinct but weak, overtone, combination, and difference bands gives rise to the appearance of a "continuum" spanning the 1–5 $\mu$m region. In actuality, it is a structured, pseudo-continuum punctuated by many weak features as suggested in the bottom rows of Figures 3 and the A1 in Appendix A. Considering the 1.6–1.7 $\mu$m region alone, there are well over 100 weak overtone and combination bands involving the CH stretching modes from only eight small PAHs (Duley 1994; Chen et al. 2019).

For highly vibrationally excited molecules, symmetry selection rules are not as rigorous as in the case of absorption because the symmetry elements of the molecule in its ground state molecular structure are no longer as well defined. Apart from expecting the intensity of an individual overtone or combination band progression to drop with decreasing wavelength, it is not possible to predict the relative intensities of these weak bands. For polyatomic molecules such as PAHs, irregular deviations from a smooth intensity drop can occur, frequently arising from perturbations induced by resonances between vibrational states (Herzberg 1968). Given the high density of vibrational states across the entire 1–5 $\mu$m region for typical interstellar PAHs, Fermi, Darling–Dennison, and vibrational Coriolis resonances will abound. Resonances between IR active states with IR dark states significantly impact peak intensities and positions. These complex interactions significantly influence the spectra, yet they are poorly understood for highly vibrationally excited large species such as interstellar PAHs.

It is likely that the bulk of the near-IR, unresolved, continuum emission from 1.25 to 4.8 $\mu$m reported by Sellgren et al. (1983), the 4.2–5.5 $\mu$m spectra from Peeters et al. (2004) shown in Figure 1, and the 2–5 $\mu$m weakly structured emission measured with AKARI toward many objects by Onaka et al. (2014) and Doney et al. (2016) is produced by the PAH transitions described here. The weakening of PAH overtone and combination bands with decreasing wavelength is likely to be responsible for the slow drop in the intensity from 5 to 1 $\mu$m in the spectra of objects that show PAH emission. Pioneering experiments measuring the IR emission from vibrationally excited small PAHs have detected weak, continuum-like emission extending well beyond the CH stretching band at 3.3 $\mu$m (Williams \& Leone 1995; Kim \& Saykally 2002; Brenner \& Barker 1992; Schlemmer et al. 1994).

The positions and relative intensities of the weak features in the 1–5 $\mu$m region are the keys to teasing out information regarding PAH structure, composition, charge, and size that is entangled in the 5–15 $\mu$m spectra. For example, the peak of the 6.2 $\mu$m band shifts from 6.22 to 6.30 $\mu$m and shows significant profile variations for many objects (e.g., Peeters et al. 2002). While variations in PAH charge (+, 0, -) or composition (PAHs with and without internal N) for the different objects are among the leading hypotheses to account for the shifts (Peeters et al. 2002; Hudgins et al. 2005), severe band blending with emission from hot bands and other transitions makes it impossible to discriminate between the various possibilities on the basis of the 6.2 $\mu$m peak position and profile alone. Measuring the overtone and combination bands involving the 6.2 $\mu$m transition directly overcomes this limitation. This is because the anharmonic vibrational energy levels for PAHs in different charge states are different from those of pure PAHs and PAHs containing N. Knowing where the overtones of the 6.2 $\mu$m band fall will enable one to separate the influence of different charge states from that of N incorporation on peak position and assess the relative contributions of both. Similar constraints emerge from knowing the precise positions of the overtones involving the severely overlapping features that comprise the 7.7 $\mu$m feature, the 8.6 $\mu$m feature, and so on. See Boersma et al. (2009) for an example of how measuring the weak PAH overtone and combination bands between 5 and 6 $\mu$m that primarily involve the strong CH-bending modes between 11 and 14 $\mu$m make it possible to extract information about interstellar PAH size (large vs. small) and structure (regular vs. compact) that is inaccessible by fundamental band analysis alone.

\section{Conclusions}

The purpose of this Letter is to draw attention to the important PAH bands predicted in the 1–5 $\mu$m region. Because of its low intensity, this region remains largely unexplored for objects showing the strong 5–15 $\mu$m PAH features. Although much weaker than the well-known PAH bands, the peak positions and relative intensities of these expected features contain information about the interstellar PAH population that is inaccessible through the strong 5–15 $\mu$m PAH features. Since the PAH features that dominate the 2.8–3.7 and 5–15 $\mu$m spectroscopic regions will be an important part of many, if not most, of the spectra JWST returns, it is critically important to fully tap the complementary information contained in the 1–5 $\mu$m region summarized below.

The positions and intensities of the deuterated PAH emission features between 4.3 to 4.5, and 4.5 to 4.8 $\mu$m as well as the CD overtone bands in objects showing the familiar 5–15 $\mu$m PAH emission features have the potential to:

\begin{enumerate}
  \item Quantify the amount of cosmological D sequestered in PAHs.
  \item Quantify the PAH D/H ratio.
  \item Show the deuterium distribution between the aliphatic and aromatic subcomponents of the PAH family in different object types.
  \item Delineate key stages in deuterated PAH formation and evolution on an object-by-object basis and within extended objects.
\end{enumerate}

\noindent
If detected, the positions and intensities of cyano-PAH bands lying between 4.46-4.51 $\mu$m, as well as the -C$\equiv$N stretch overtones in objects showing the familiar PAH bands in the 5-15 $\mu$m range have the potential to:
\begin{enumerate}
  \item Determine whether nitrogenated PAHs are part of the cosmic PAH population.
  \item Quantify the N/C-ratio in the PAH population.
  \item Track the growth and loss of PAH-nitriles through different stages of the stellar lifecycle,
        revealing previously hidden, critical details about PAH growth and fragmentation pathways.
\end{enumerate}
\noindent
The positions and intensities of PAH overtone and combination bands are the keys to access fundamental information about PAH structure, composition, charge, and size that is entangled in the 5-15 $\mu$m spectra.

Additionally, the measurement of these astronomical features will motivate computational and experimental studies of the overtone and combination transitions in large PAHs, the fundamental and overtone -C$\equiv$N transitions in cyano-PAHs, and the overtones of the aromatic and aliphatic CD transitions in deuterated PAHs, similar to the vigorous work that followed the discovery of the PAH emission features nearly fifty years ago.\\

\noindent
L.J.A., C.B., and J.D.B. gratefully acknowledge NASA SMD support from a directed Work Package titled: "Laboratory Astrophysics—The NASA Ames PAH IR Spectroscopic Database". T.J.L. gratefully acknowledges support from NASA grants 17-APRA17-0051, 18-APRA18-0013, and 18-2XRP18 2-0046. L.J.A. and J.D.B. are both thankful for an appointment at NASA Ames Research Center through the Bay Area Environmental Research Institute (80NSSC17M0014). C.B. is appreciative of an appointment at NASA Ames Research Center through the San José State University Research Foundation (NNX17AJ88A).

\vskip1cm

{\centering
  \bf ORCID iDs

}

\vskip0.2cm

\noindent
L. J. Allamandola https:/orcid.org/0000-0002-6049-4079 \\
C. Boersma https:/orcid.org/0000-0002-4836-217X\\
T. J. Lee https:/orcid.org/0000-0002-2598-2237\\
J. D. Bregman https:/orcid.org/0000-0002-1440-5362\\
P. Temi https:/orcid.org/0000-0002-8341-342X

\vskip1cm

{\centering
  \bf References

}

Agundez, M., Cernicharo, J., Quintana-Lacaci, G., et al. 2017, A\&A, 601, A4 \\
Bakes, E. L. O., \& Tielens, A. G. G. M. 1994, ApJ, 427, 822\\
Bernstein, M., Sandford, S., Allamandola, L., et al. 1999, Sci, 283, 1135 \\
Bernstein, M. P., Elsila, J. E., Dworkin, J. P., et al. 2002, ApJ, 576, 1115 \\
Boersma, C., Bauschlicher, C. W., Ricca, A., et al. 2011, ApJ, 729, 64 \\
Boersma, C., Mattioda, A. M., Bauschlicher, C., et al. 2009, ApJ, 690, 1208 \\
Boschman, L., Cazaux, S., Spaans, M., et al. 2015, A\&A, 579, 72\\
Brenner, J. D., \& Barker, J. R. 1992, ApJ, 388, L39\\
Cami, J., Bernard-Salas, J., \& Peeters, E. 2010, Sci, 329, 1180\\
Cernicharo, J., Guelin, M., \& Kahane, C. 2000, A\&AS, 142, 181 \\
Cernicharo, J., Heras, A. M., Tielens, A. G. G. M., et al. 2001, ApJL, 546, L123\\
Chen, T., Luo, Y., \& Li, A. 2019, A\&A, 632, A71\\
Doney, K. D., Candian, A., Mori, T., Onaka, T., \& Tielens, A. G. G. M. 2016, A\&A, 586, A65\\
Draine, B. T. 2004, in in Origin and Evolution of the Elements, from the Carnegie Observatories Centennial Symposia, ed. A. McWilliam \&
M. Rauch (Cambridge: Cambridge Univ. Press), 317\\
Duley, W. W. 1994, ApJ, 429, L91\\
Dworkin, J. P., Gillette, J. S., Bernstein, M. P., et al. 2004, AdSpR, 33, 67 \\
Fonfria, J., Cernicharo, J., Richter, M. J., \& Lacy, J. H. 2008, ApJ, 673, 445 \\
Geballe, T., Joblin, C., d’Hendecourt, L., et al. 1994, ApJ, 434, L15 \\
Greenberg, J. M., Gillette, J. S., Munoz Caro, G. M., et al. 2000, ApJ, 531, L71 \\
Hamid, A. M., Bera, P. B., Lee, T. J., et al. 2014, J. Phys. Chem. Lett., 5, 3392 \\
Herzberg, G. 1968, Molecular Spectra and Molecular Structure II. IR and
Raman Spectroscopy of Polyatomic Molecules (3rd ed.; Princeton, NJ: D.
Van Nostrand Company Inc.)\\
Hoy, A. R., Mills, I. M., \& Strey, G. 1972, MolPh, 24, 1265\\
Hudgins, D., Bauschlicher, C., \& Allamandola, L. J. 2005, ApJ, 632, 316 \\
Hudgins, D., Bauschlicher, C., \& Sandford, S. A. 2004, ApJ, 614, 770 \\
Inostroza, N., Fortenberry, R. C., Huang, X., \& Lee, T. J. 2013, ApJ, 778, 160 \\
Kaiser, R. I., Parker, D. S. N., \& Mebel, A. M. 2015, ARPC, 66, 43 \\
Kalvans, J., \& Shmeld, I. 2018, arXiv:1803.05731\\
Kamp, I., \& Dullemond, C. P. 2004, ApJ, 615, 991\\
Kim, H., \& Saykally, R. J. 2002, ApJSS, 143, 455\\
Lee, T. J., Huang, X., \& Dateo, C. E. 2009, MolPh, 107, 1139\\
Lee, T. J., Martin, J. M. L., Dateo, C. E., \& Taylor, P. R. 1995, JPhCh, 99, 15858\\
Li, A. 2020, NatAs, 4, 339\\
Li, A., \& Draine, B. T. 2001, ApJ, 554, 778\\
Linsky, J., Draine, B. T., Moos, H. W., et al. 2006, ApJ, 647, 1106\\
Mackie, C. J., Candian, A., Huang, X., et al. 2018a, PCCP, 20, 1189 \\
Mackie, C. J., Chen, T., Candian, A., Lee, T. J., \& Tielens, A. G. G. M. 2018b, JChPh, 149, 134302\\
Maltseva, E., Mackie, C. J., Candian, A., et al. 2018, A\&A, 610, A65 \\
Maltseva, E., Petrignani, A., Candian, A., et al. 2015, ApJ, 814, 23 \\
McGuire, B. A., Burkhardt, A. M., Sergei, K., et al. 2018, Sci, 359, 202 \\
McGuire, B. A., Loomis, R. A., Burkhardt, A. M., et al. 2021, Sci, 371, 1265 \\
Onaka, T., Mori, T. I., Sakon, I., et al. 2014, ApJ, 780, 114\\
Parker, D. S. N., \& Kaiser, R. I. 2017, Chem. Soc. Rev., 46, 452\\
Peeters, E., Allamandola, L. J., Bauschlicher, C., et al. 2004, ApJ, 604, 252 \\
Peeters, E., Hony, H., van Kerckhoven, C., et al. 2002, A\&A, 390, 1089 \\
Reeves, H., Audoze, J., Fowler, W. A., \& Schramm, D. N. 1973, ApJ, 179, 909\\
Sandford, S. A., Bernstein, M. P., Allamandola, L. J., Gillette, J. S., \& Zare, R. N. 2000, ApJ, 538, 691\\
Santoro, G., Martinez, L., Lauwaet, K., et al. 2020, ApJ, 895, 97 \\
Schlemmer, S., Cook, D. J., Harrison, J. A., et al. 1994, Sci, 265, 1686 \\
Sellgren, K., Werner, M. W., \& Dinerstein, H. K. 1983, ApJL, 271, L13 \\
Sellgren, K., Werner, M. W., Ingalls, J. G., et al. 2010, ApJL, 722, L54 \\
Silverstein, R. M., \& Bassler, G. C. 1967, Spectrometric Identification of Organic Compounds (New York: Wiley)\\
Smith, J. D. T., Draine, B. T., Dale, D. A., et al. 2007, ApJ, 656, 770\\
Soliman, A., Attah, I. K., Hamid, A. M., \& El-Shall, S. 2015, IJMSp, 377, 139 \\
Thrower, J. D., Jorgensen, B., Friis, E. E., et al. 2012, ApJ, 752, 3\\
Tielens, A. G. G. M. 2008, ARA\&A, 46, 289\\
Verstraete, L. 2011, EAS, 46, 415\\
Watson, J. K. G. 1977, in in Vibrational Spectra and Structure, 6 ed. J. R. Durig (Amsterdam: Elsevier), 1\\
Weingartner, J. C., \& Draine, B. T. 2001, ApJ, 563, 842\\
Wiersma, S. D., Candian, A., Bakker, J. M., et al. 2020, A\&A, 635, A9 \\
Williams, R. M., \& Leone, S. R. 1995, ApJ, 443, 675\\


\renewcommand\thefigure{\thesection\arabic{figure}}
\renewcommand\thetable{\thesection\arabic{table}}
\setcounter{figure}{0}
\setcounter{table}{0}
\appendix

\section{PAH Combination Bands}

The treatment for determining combination band positions (scaling harmonic frequencies) is described below, whereas the different treatment for determining overtone band positions (computing anharmonic frequencies) is described in Appendix B. The latter describes an approach in which the energy levels for a particular vibrational mode get closer together when moving to higher energy levels, as shown in the left panel of Figure 1 and as occurs in real systems. However, as discussed further at the end of Appendix B, the approximation used in Appendix B cannot be applied to combination bands. Hence, the scaled harmonic approximation described below is used for combination bands.

Figure A1 below illustrates very few of the many possible combination bands that fall across the 1–5 $\mu$m region. These are scaled harmonically, based on the canonical peak positions of the prominent PAH emission features between 5 and 15 $\mu$m. Keep in mind that overtones and combinations involving the many other, longer-wavelength PAH fundamental modes (Boersma et al. 2011) will also contribute to the emission in this region.

\begin{figure*}[ht!]
  \begin{center}
    \hskip-0.0cm
    \includegraphics[width=13.8cm]{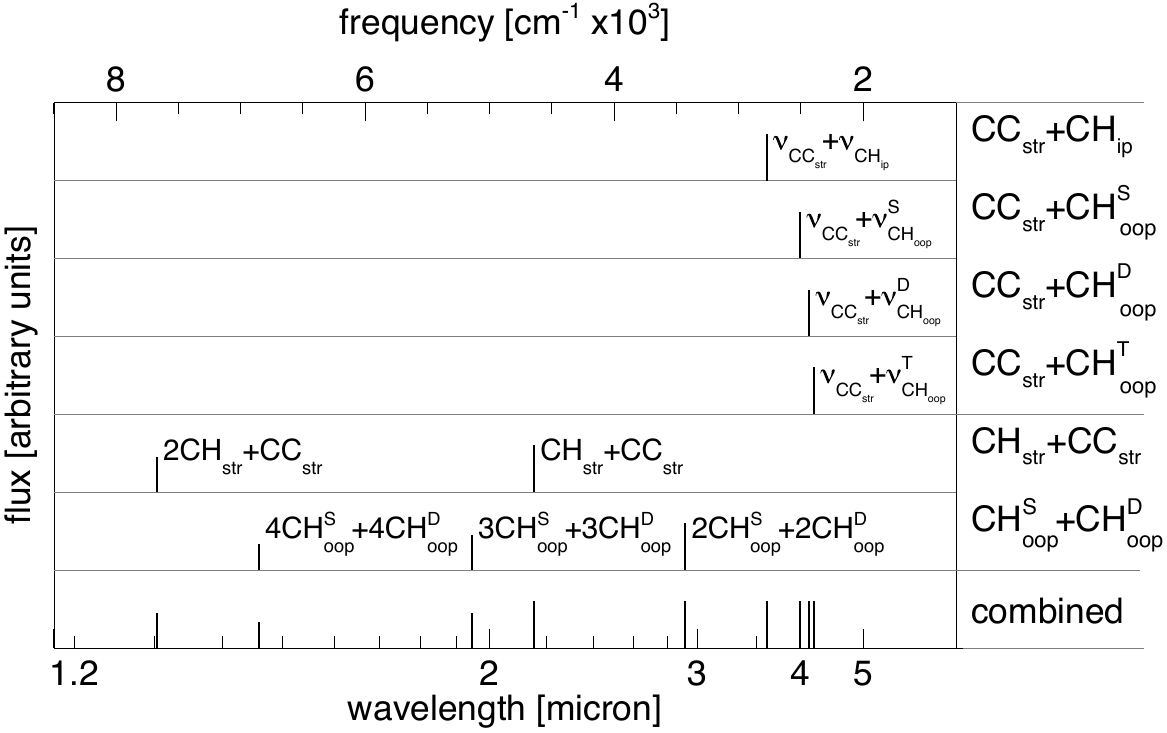}
    \caption{
      Schematic representation of some of the scaled harmonic positions of the PAH combination bands that fall in the 1 - 5 $\mu$m region. Band intensities are shown to fall by 25\% with each step. Vibrational mode types: str- stretch; ip- in plane CH bend; oop- out of plane CH bend; S - solo hydrogen; D - duo hydrogen; T - trio hydrogen. Vibrational mode types: str - stretch; ip-in plane CH bend; oop- out of plane CH bend; S - solo hydrogen; D - duo hydrogen; T - trio hydrogen.}
    \label{fg:appendix}
  \end{center}
\end{figure*}

\section{Determining Anharmonicities}

Over the last few years, vibrational second-order perturbation theory (VPT2; Hoy et al. 1972; Watson 1977) has been used to determine the anharmonic vibrational spectra of PAH molecules and PAH derivative molecules (Mackie et al. 2018a and Maltseva et al. 2018) with significant success provided that resonances are treated by a polyad. In fact, Mackie et al. (2018b) have been able to use VPT2 in conjunction with statistical models to simulate fully anharmonic cascade emission spectra of these PAH molecules. VPT2 is ideally suited to studying the anharmonicity of PAH molecules (i) because the PAH molecules are fairly rigid (i.e., no large-amplitude motions) and (ii) because of their size, even after absorbing a 12 eV photon and then redistributing that energy around the various vibrational modes, the average v quantum number for any one vibrational mode in the mid-IR is less than 1. As such, we have used VPT2 to approximate the anharmonicity of the overtone wavelengths of the $\Delta$v=1$\rightarrow$0, 2$\rightarrow$0 and 3$\rightarrow$0 CD and C$\equiv$N stretching bands listed in Table 1 and shown in Figure 2. The vibrational energy levels for an asymmetric top computed from VPT2 are given by:

\begin{equation} \label{eq1}
  \begin{split}
    E(\textrm{v}) & = \sum^N_{k=1} \omega_k \left( \textrm{v}_k + \frac {1}{2} \right)\\
    &  +  \sum^N_{k \le l} X_{kl}  \left( \textrm{v}_k + \frac {1}{2} \right) \left( \textrm{v}_l + \frac {1}{2} \right)
  \end{split}
\end{equation}

where v$_k$ is the vibrational quantum number for normal mode k, N is the total number of vibrational modes, $\omega_k$ is the harmonic frequency for normal mode k, and X$_{kl}$ is the anharmonic constant connecting normal modes k and l.  Examination of Eq. (1) shows that the zero-point vibrational energy level is given by:\\

\begin{equation} \label{eq2}
  E(\textrm{v})  = \frac{1}{2}  \sum^N_{k=1} \omega_k + \frac{1}{4}  \sum^N_{k \le l} X_{kl}
\end{equation}

(Note that this is not the VPT2 zero-point vibrational energy (ZPVE) because there is an anharmonic constant term that is needed for thermochemistry applications, but for transition energy calculations, this constant term cancels out.)  The VPT2 fundamental vibrational frequency for normal mode k is given by $\nu_k$=E(v$_k$=1)-E(v$_k$=0),
where it is understood that the quantum numbers v for all of the other normal modes are zero for both energy levels.  As discussed by Mackie et al. (2018b) in modeling the cascade emission spectra of PAH molecules, there will be spectator modes occupied, but the change in the transition energy generally will be small as the coupling between the low-energy bending vibrational modes and the high-energy stretching modes discussed here will be small. The small changes would tend toward increasing the anharmonic correction, or in other words making the transition energy slightly smaller, but since the anharmonic correction for the C–D aromatic and aliphatic stretches is larger than the -C$\equiv$N stretch having occupied spectator modes would have the effect of increasing the separation between the C and D and -C$\equiv$N overtone transition energies. For purposes of this study, it is a reasonable approximation to assume that there are no spectator modes populated.

Working this out from Eq. (1), one obtains\\

\begin{equation} \label{eq3}
  \nu_k  = \omega_k + 2 X_{kk} + \frac{1}{2} \sum^N_{k<l} X_{kl}
\end{equation}

One can generalize the energy difference between successive levels to the following\\

\begin{equation} \label{eq4}
  (\textrm{v}_k  + 1) \nu_k - \textrm{v}_k \nu_k = \omega_k + 2 X_{kk} (\textrm{v}_k  + 1) + \frac{1}{2} \sum^N_{k<l} X_{kl}
\end{equation}

As noted, in Equation (3) it is assumed that the v quantum numbers for all other normal modes is zero. Equation (3) will change slightly if there are spectator modes included (a spectator mode is defined as one where its quantum number v has the same non-zero value in both vibrational states being considered). This small change involves only the X$_{kl}$ constants for asymmetric tops. Further, Equation (3) is also correct for symmetric and spherical top molecules provided there are no spectator modes included, the normal mode k is a non-degenerate vibrational mode, and any degenerate vibrational modes are summed over each component in the last terms (i.e., treat each component of a degenerate vibrational frequency separately).

Another way to examine these changes is the overtone frequencies themselves, i.e. E(v$_k$=1)-E(v$_k$=0), which we will write as $n\nu_k$ . Starting from Equation (1), we obtain the following general formula\\

\begin{equation} \label{eq5}
  \begin{split}
    E(\textrm{v}_k=n) - E(\textrm{v}_k=0)& = n\omega_k + X_{kk} \left[ \left(n+\frac{1}{2}\right)^2 - \frac{1}{4}\right] \\
    &  +  \sum^N_{k < l} X_{kl}   \left[ \left(n+\frac{1}{2}\right) \left(\frac{1}{2} \right)- \frac{1}{4}\right]
  \end{split}
\end{equation}

where again it is assumed that there are no spectator modes populated.

\begin{table}[ht]
  \small
  \caption{Anharmonic Corrections for the Fundamental and Overtones of C$\equiv$N Stretches, $cm^{-1}$}
  \begin{center}
    \begin{tabular}{lccccc}
      \hline
      \hline
      Molecule & $\omega$  & $\nu$     & $\nu-\omega$ & $2\nu-2\omega$ & $3\nu-3\omega$ \\
               & $cm^{-1}$ & $cm^{-1}$ & $cm^{-1}$    & $cm^{-1}$      & $cm^{-1}$      \\
      \hline
      HCN      & 2109.4    & 2079.4    & -30.0        & -79.9          & -149.9         \\
      \hline
      FCN      & 2350      & 2312      & -38          & -103           & -192           \\
      \hline
      ClCN     & 2242      & 2209      & -33          & -91            & -173           \\
      \hline
      CN       & 2067.7    & 2041.8    & -25.9        & -77.6          & -155.2         \\
      \hline
      DCCN     & 2061.4    & 2042.3    & -19.1        & -101.8         & -196.2         \\
      \hline
    \end{tabular}
    \label{tbl:B1}
  \end{center}
\end{table}

The -C$\equiv$N and C–D anharmonic corrections incorporated in the results in Table 1 and Figure 2 were obtained using successive applications of Equation (4). The -C$\equiv$N anharmonic corrections were obtained by examining the VPT2 fundamental vibrational frequencies of various small -C$\equiv$N containing molecules determined using the singles and doubles coupled-cluster method with a perturbational treatment of triple excitations, denoted CCSD(T). These include HCN, FCN, ClCN, neutral CN, and finally DCCN. First, we note that the fundamental for -C$\equiv$N attached to a PAH molecule is found to be in the 2242–2222 $cm^{-1}$ range (Silverstein \& Bassler 1967). Of the molecules examined, the -C$\equiv$N stretch for ClCN is closest to this range occurring at 2209 $cm^{-1}$ (Lee et al. 1995), hence the anharmonic corrections for ClCN were chosen to determine the $\Delta$v=2$\rightarrow$0 and 3$\rightarrow$0 transition energies. This leads to a range of 4426–4406 $cm^{-1}$ (2.259–2.270 $\mu$m) for $\Delta$v=2$\rightarrow$0 and 6586–6566 $cm^{-1}$ (1.518–1.523 $\mu$m) for $\Delta$v=3$\rightarrow$0.

For the aromatic -C–D stretch, the fundamental range is 2326–2222 $cm^{-1}$ (Hudgins et al. 2004) and the molecules examined (using the CCSD(T) method) include c-C3 HD, c-C3 D2, and c-C3 H2 D +. In this case, the antisymmetric C–D stretch in c-C3 D2 is closest to this range at 2343 $cm^{-1}$ (Lee et al. 2009). Using the procedure described above, we obtain a range of 4517–4621 $cm^{-1}$ (2.214–2.164 $\mu$m) for $\Delta$v=2$\rightarrow$0, and 6781–6885 $cm^{-1}$ (1.475–1.452 $\mu$m) for $\Delta$v=3$\rightarrow$0.

For the aliphatic -C–D stretch, the fundamental range is 2203-2105 $cm^{-1}$ (Hudgins et al. 2004) and the molecules whose CCSD(T) VPT2 fundamentals were examined include CD4 and DCCN. In this case, the -C–D fundamental for DCCN is closest to this range at 2187 $cm^{-1}$ (Inostroza et al. 2013). Using the procedure described above, we obtain a range of 4222–4320 $cm^{-1}$ (2.369–2.315 $\mu$m) for $\Delta$v=2$\rightarrow$0, and a range of 6268–6366 $cm^{-1}$ (1.595–1.571 $\mu$m) for $\Delta$v=3$\rightarrow$0.

The procedure described here for determining the frequency or wavelength range for fundamental and overtone bands is well established for specific vibrational modes such as the -C$\equiv$N and the aromatic and aliphatic -C–D stretches. In fact, this procedure is most applicable to stretching vibrational modes. That is, the anharmonic corrections for the fundamental and overtone bands fall within a narrow range for well-defined stretching vibrational modes even though individual anharmonic constants are not transferrable from one molecule to another. As an example, Table B1 contains the CCSD(T) harmonic frequency ($\omega$), fundamental frequency ($\nu$), and the anharmonic corrections for the fundamental ($\nu-\omega$), and first ($2\nu-2\omega$) and second ($3\nu-3\omega$) overtone bands for the CN stretching frequency in HCN, FCN, ClCN, CN, and DCCN, the molecules mentioned above.

The largest variation in CN stretching frequencies occurs for the harmonic frequencies, which is due to the chemical nature of the molecule in which the -C$\equiv$N bond occurs. Even though there is a 289 $cm^{-1}$ range for the harmonic frequencies (2350–2061 $cm^{-1}$), the range for the anharmonic correction to the fundamental, ($\nu-\omega$), is only $\sim$19 $cm^{-1}$ (i.e., -19 to -38 $cm^{-1}$). The range for the anharmonic corrections of the first and second overtone bands is likewise narrow. Since the anharmonic corrections for the first and second overtone bands become sequentially larger than the fundamental the energy levels with increasing vibrational level (v), the levels draw closer together, consistent with Figure 1 and reality. To put this in an observational context, a +19 $cm^{-1}$ shift at 4.50 $\mu$m (2222 $cm^{-1}$) adjusts to 4.48 $\mu$m (2231 $cm^{-1}$).

Unfortunately, this procedure does not apply to combination bands for two reasons. One, for a combination band, all the molecules examined would need to include both vibrational modes that make up the combination band and the coupling between these vibrational modes would need to be the same for each molecule. This is unlikely to be the case for most molecules. Two, as indicated previously, the narrow range in anharmonic corrections occurs for well-defined stretching vibrational modes (i.e., tight chemical bonds, so no large-amplitude motions). This will not necessarily occur for other types of vibrational modes such as bends, out-of-plane bends, etc. It will almost certainly be the case that the combination bands will be composed of vibrational modes that are not stretches (see Figure 1 Appendix A), and thus the anharmonic corrections will be more system dependent. Therefore, the procedure described in Appendix A (simply scaling harmonic frequencies) was used for combination bands.

\end{document}